\documentclass[aps,prl,twocolumn,groupedaddress,showpacs,floatfix]{revtex4}
\usepackage{graphicx}
\usepackage{amssymb}
\usepackage{xspace} 
\usepackage{ulem}
\usepackage[breaklinks=true,colorlinks=true,linkcolor=blue]{hyperref}

\begin{document}

\newcommand{\qweak}{$Q_{\rm weak}$\xspace}
\newcommand{\fp}{Fabry-P\'{e}rot\xspace}


\title{ Precision Electron-Beam Polarimetry at 1 GeV
  Using Diamond Microstrip Detectors}
\author{A.~Narayan$^1$}
\author{D.~Jones$^2$}
\author{J.~C.~Cornejo$^3$}
\author{M.~M.~Dalton$^{2,4}$}
\author{W.~Deconinck$^3$}
\author{D.~Dutta$^1$\footnote{corresponding author, d.dutta@msstate.edu}}
\author{D.~Gaskell$^4$}
\author{J.~W.~Martin$^5$}  
\author{K.D.~Paschke$^2$}
\author{V.~Tvaskis$^{5,6}$}
\author{A.~Asaturyan$^7$}
\author{J. Benesch$^4$}
\author{G.~Cates$^2$}
\author{B.~S.~Cavness$^8$}
\author{L.~A.~Dillon-Townes$^4$}
\author{G.~Hays$^4$}
\author{E.~Ihloff$^{9}$}
\author{R.~Jones$^{10}$}
\author{P.~M.~King$^{11}$}
\author{S.~Kowalski$^{12}$}
\author{L.~Kurchaninov$^{13}$} 
\author{L.~Lee$^{13}$}
\author{A.~McCreary$^{14}$}
\author{M.~McDonald$^5$}
\author{A.~Micherdzinska$^5$}
\author{A.~Mkrtchyan$^7$}
\author{H.~Mkrtchyan$^7$}
\author{V.~Nelyubin$^2$}
\author{S.~Page$^6$}
\author{W.~D.~Ramsay$^{13}$}
\author{P.~Solvignon$^4$}
\author{D.~Storey$^5$}
\author{A.~Tobias$^2$}
\author{E.~Urban$^{15}$}
\author{C.~Vidal$^{9}$}
\author{B. Waidyawansa$^{11}$}
\author{P.~Wang$^6$}
\author{S.~Zhamkotchyan$^7$}
\affiliation{$^1$Mississippi State University, Mississippi State, MS 39762, USA}
\affiliation{$^2$University of Virginia, Charlottesville, VA 22904, USA} 
\affiliation{$^3$College of William and Mary, Williamsburg, VA 23187, USA}
\affiliation{$^4$Thomas Jefferson National Accelerator Facility, Newport News, VA 23606, USA}
\affiliation{$^5$University of Winnipeg, Winnipeg, MB R3B~2E9, Canada}
\affiliation{$^6$University of Manitoba, Winnipeg, MB R3T~2N2, Canada}
\affiliation{$^7$Yerevan Physics Institute, Yerevan, 375036, Armenia}
\affiliation{$^8$Angelo State University, San Angelo, TX 76903, USA}
\affiliation{$^9$MIT Bates Linear Accelerator Center, Middleton, MA 01949, USA}
\affiliation{$^{10}$University of Connecticut, Storrs, CT 06269, USA}
\affiliation{$^{11}$Ohio University, Athens, OH 45701}
\affiliation{$^{12}$Massachusetts Institute of Technology, Cambridge, MA 02139, USA}
\affiliation{$^{13}$TRIUMF, Vancouver, BC V6T~2A3, Canada}
\affiliation{$^{14}$University of Pittsburgh, Pittsburgh, PA 15260, USA}
\affiliation{$^{15}$Hendrix College, Conway, AR 72032, USA} 

\begin{abstract}
We report on the highest precision yet achieved in the measurement of the polarization of a
low-energy, $\mathcal{O}$(1  GeV), continuous wave (CW) electron beam, accomplished using a new polarimeter
based on electron-photon scattering, in Hall~C at Jefferson Lab.  
A number of technical innovations were necessary, including a novel method for precise
control of the laser polarization in a cavity and a novel diamond microstrip detector
which was able to capture most of the spectrum of scattered electrons. The data analysis technique 
exploited track finding, the high granularity of the detector and its
large acceptance. The polarization of the $180~\mu$A, $1.16$~GeV electron beam was measured with a
statistical precision of $<$~1\% per hour and a systematic uncertainty of 0.59\%.
This exceeds the level of precision required by the \qweak experiment, a measurement
of the weak vector charge of the proton. Proposed future low-energy experiments
require polarization uncertainty $<$~0.4\%, and this
result represents an important demonstration of that possibility.  This measurement is the first use 
of diamond detectors for particle tracking in an experiment. It demonstrates the stable operation of a diamond based tracking detector in a high radiation environment, for two years.

\end{abstract}


\maketitle

\section{Introduction}
\label{sec:intro}

High-precision physics experiments using polarized electron beams rely on accurate knowledge
of beam polarization to achieve their ever improving precision. A parity-violating electron-scattering experiment in Hall~C at Jefferson Lab (JLab), known as the \qweak
experiment, is the most recent example~\cite{qweakprl,qweaknim}.   The \qweak experiment
aims to test the Standard Model of particle physics by providing a first precision
measurement of the weak vector charge of the proton, from which the weak mixing angle will
be extracted with the highest precision away from the Z$^0$ pole. With the \qweak
experiment proposed to obtain a statistical precision of 2.1\% on the parity-violating asymmetry,
the uncertainty goal for beam polarimetry was 1\%. Two future precision
Standard Model tests at JLab, SOLID and MOLLER, have far more stringent polarimetry requirements
of 0.4\%~\cite{MOLLERExp,SOLID}.


In order to meet the high-precision requirement of the \qweak experiment, a new polarimeter based on electron-photon scattering (Compton scattering) was constructed
in experimental Hall~C~\cite{qweaknim,narayanthesis}. This polarimeter could be operated without disrupting the electron beam, allowing
for continuous polarization measurement during the \qweak experiment.  An existing polarimeter in Hall C, using a magnetized iron foil
target to measure polarized $e^{-}e^{-}$ scattering (M\o ller scattering), has previously reported a polarization measurement significantly
better than 1\%~\cite{moller,Magee}.  However, the M\o ller measurement is destructive to the polarized beam and requires reduced beam
current, and therefore the results must be extrapolated in beam current and interpolated in time between the dedicated measurements.  

\begin{figure*}[t!]
\includegraphics*[height=16cm,angle=-90]{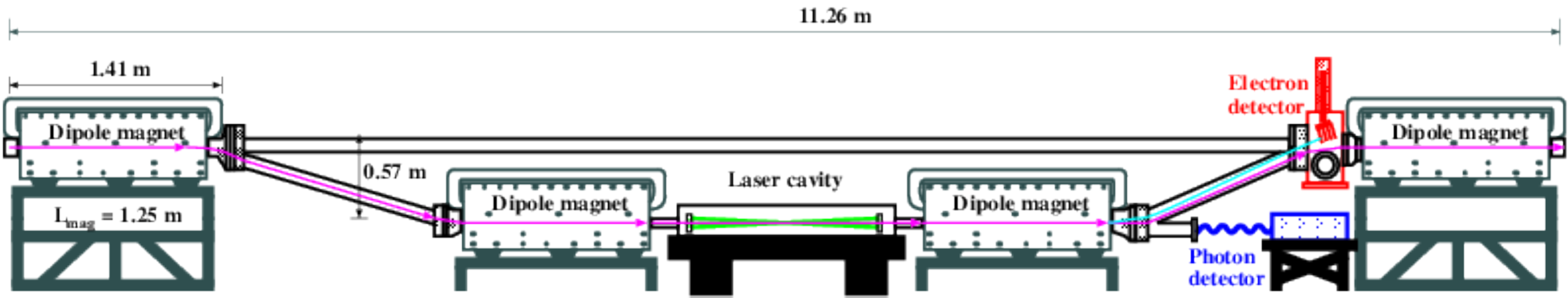}
\caption[]{Schematic diagram of the JLab Hall~C Compton polarimeter. Four identical dipole magnets form a magnetic chicane that displaces
  the 1.16~GeV electron beam vertically downward by 0.57~m. 
  An external low-gain \fp laser cavity provides a high intensity ($\sim 1.7$~kW) beam of circularly polarized green (532~nm) photons. The
  laser light is focused at the interaction region ($\sigma_{\rm {waist}} \sim$ 90~$\mu$m), and it is larger than the electron beam
  envelope ($\sigma_{\rm {x/y}} \sim$ 40~$\mu$m).  The photon detector was not used for these results.} 
\label{fig:fig1}
\end{figure*}

In this report we present the first measurement of electron beam polarization with the
new Hall~C Compton polarimeter, with the best precision ever achieved in this energy range (0.6\%), and we directly compare the result with
the Hall C M\o ller polarimeter.  
With each polarimeter reporting precision better than 1\%, a direct comparison of the two independent measurements provides  a valuable
cross-check of electron-beam polarimetry techniques. These results also suggest that the rigorous demands of future experiments can be met.

Compton polarimetry is an established technique~\cite{SPEAR,LEP, Hera, Nikhef, LPOL, Bates, HallA, Escoffier, Friend, SLD, happex2} which involves measuring a known
QED double-spin asymmetry in electron scattering from a photon beam of known polarization. The scattering asymmetry varies with the fraction
of electron beam energy transferred to the scattered photon, with the maximum asymmetry occurring at the kinematic limit for maximum
backscattered photon energy.
The Compton-scattered electrons and photons can be independently measured and analyzed to determine the polarization of the electron beam.
Most Compton polarization measurements have primarily analyzed the backscattered photons~\cite{SPEAR,LEP, Hera, Nikhef, Bates, HallA, Escoffier, Friend}
and reliance on electron measurements has been less common~\cite{SLD,happex2}. 
Both the maximum scattering asymmetry and the maximum fraction of beam energy transferred to the photon increase quadratically with beam energy.
For this reason, Compton scattering measurements are significantly more difficult at low beam energies.

The SLC Large Detector (SLD) Compton polarimeter at the Stanford Linear Collider (SLC)~\cite{SLD}
detected scattered electrons in a segmented gas Cerenkov detector with a reported precision of
0.5\%---the only Compton polarimetry measurement more precise than this work.   
Operating at lower energies, the Compton polarimeter in Hall~A at Jefferson Lab has reported a precision of $\sim$1\% by detecting the Compton scattered electrons in a silicon microstrip detector~\cite{happex2} at a beam energy of 3~GeV and, in separate measurements, by integrating the total power of Compton scattered photons in a total-absorption Gadolinium oxyorthosilicate (GSO) calorimeter~\cite{Parno} at 1-3~GeV~\cite{Friend,PREX}. 

The \qweak measurement presented new challenges to this established polarimetry technology. 
The very precise SLD result was achieved with a 532~nm laser at a beam energy of 46.5~GeV, providing a maximum asymmetry $A_{\rm exp} \sim 0.75$ and a maximum photon energy of almost 60\% of the electron beam energy.  At the relatively low energy (1.16~GeV) of the electron
beam for the \qweak experiment, the maximum Compton asymmetry$A_{\rm exp}~\sim~0.04$ is significantly smaller, and only 5\% of the electron
energy can be transferred to the photon. The small asymmetry requires large luminosity to achieve sufficient statistical precision, while
the lower kinematic limit implies that an electron detector must be positioned close to the primary beam and have high granularity to achieve suitable resolution on the scattered electron momentum.  

The electron accelerator at Jefferson Lab operates at 1497 MHz with a beam repetition rate of 499 MHz to each of the three experimental halls and a bunch width of $\approx$ 0.5 ps. The small 2~ns
  spacing between each beam bunch implies that from the perspective of most detectors, the electron beam is
  essentially CW. In Compton 
polarimeters used in colliders (for example at the Hadron Electron Ring Accelerator (HERA) and SLD), the repetition rate of the electron beam was quite 
modest (on the order of 100s of Hz) 
and it was possible to use low average power, pulsed lasers to achieve high instantaneous scattering rates
and hence excellent background suppression. This approach was not possible at Jefferson Lab, and a CW 
laser system was required.

The desired high luminosity was achieved by storing laser photons in a \fp cavity, 
even though past measurements of the laser polarization have proven to be challenging 
in evacuated \fp cavities. An innovative technique for maximizing the laser polarization 
by analyzing the reflected light at the cavity entrance was employed during the \qweak experiment.

The high signal count rate, expected large background close to the beam, and proposed experimental run of 200 days required the
selection of radiation-hard detection systems. A diamond microstrip detector was selected for electron detection. The well-established
radiation hardness of diamond~\cite{Bauer1995,Zoeller1997} and its insensitivity to synchrotron radiation were the most important
considerations in this choice. 
While diamond microstrip detectors have been demonstrated in test beams~\cite{Borchelt,Tapper2000}, and other diamond detector
configurations have been used in beam condition monitors~\cite{tesla,babar,cdf,cms,gsi,atlas}, this is the first application of a
diamond detector in
an experiment as a particle tracking detector.

\section{The Hall~C Compton Polarimeter}
\label{sec:cpol}

A schematic of the Compton polarimeter in Hall C at JLab is shown in Fig.~\ref{fig:fig1}, and details can be found in
Ref.~\cite{qweaknim,narayanthesis}. The CW electron beam was deflected vertically by two dipole magnets to where it could interact with the
photon target. 
Circularly polarized 532-nm laser light was injected into a \fp optical cavity, in the beamline vacuum, with a gain of approximately 200.
The injection laser, a Coherent Verdi~\cite{verdi} with an output of 10~W, was locked to the cavity. 
The 0.85~m long optical cavity crossed the electron beam at 1.3$^\circ$. 

After interacting with the photon target, the electron beam was deflected back to the nominal beamline with a second pair of dipole magnets.
The Compton scattered photons passed through an aperture in the third magnet and were detected in an array of PbWO$_4$ crystals. The analysis of the detected photons were used as a crosscheck of the electron analysis. The third chicane magnet bent the primary beam by 10.1$^{\circ}$, also separating the Compton scattered electrons
from the primary beam by up to 17~mm before the fourth dipole. Here the scattered electrons were incident on the electron detector, a set of
four planes of diamond microstrip detectors.  Remote actuation allowed the detector distance to the primary
beam to be varied.  Data were taken with the innermost strip a mere 5~mm from the beam, with
routine operation at 7~mm from the beam. This range allowed the detection of most of the Compton electron spectrum, including the
zero-crossing of the asymmetry 8.5~mm from the primary beam.

\begin{figure}[hbtp!]
\includegraphics*[width=9.0cm]{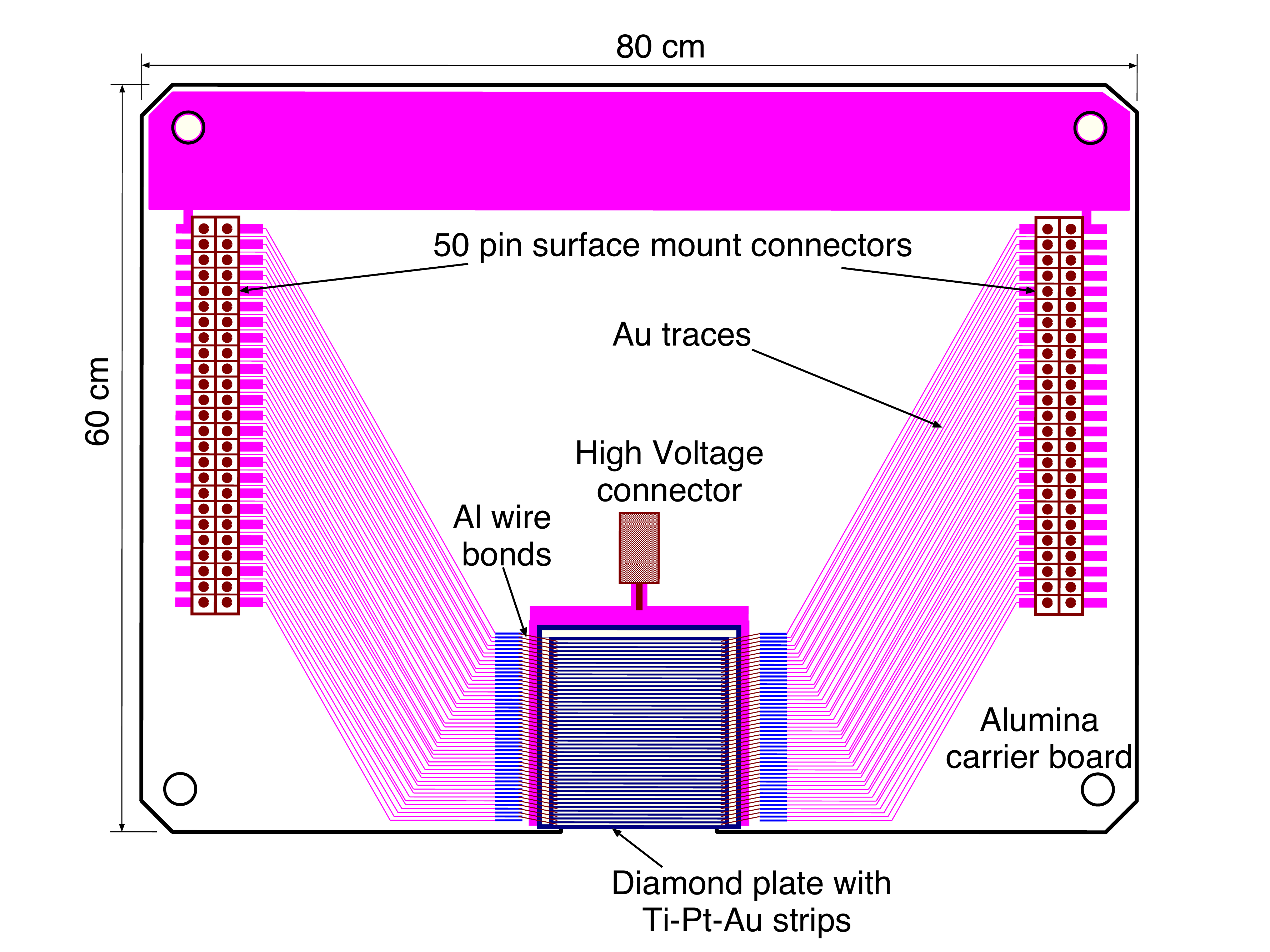}
\caption[]{A schematic diagram of the CVD diamond plate mounted on an alumina frame which forms a single detector plane. There were 96 Ti-Pt-Au
  strips deposited on the front face of the diamond plate which was attached to the frame using a silver epoxy. The strips were connected to
  Au traces on the alumina frame with aluminum wire bonds. The traces terminated on two 50 pin connectors. A high voltage (HV) bias of $\sim$ -300~V
  was applied to the back side of the diamond plate via a miniature HV connector.} 
\label{fig:edet}
\end{figure}

The electron detectors were made from 
$21\times21\times0.5$~mm$^3$ plates of synthetic diamond grown using chemical vapor deposition (CVD)~\cite{ddet}. A novel Ti-Pt-Au
metalization was used to deposit electrodes on the diamond plates. Each diamond plate has 96 horizontal metalized electrode strips
with a pitch of 200~$\mu$m (180~$\mu$m of metal and a 20~$\mu$m gap) on one side. The Compton spectrum is spread 
over 50~--~60 strips allowing a precise measurement of the shape. A schematic of a single detector plane is shown in Fig~\ref{fig:edet}.
 The strips were read out using custom low noise amplifiers and discriminators, grouped together with 48
channels in a single module~\cite{qwad}. The detector signal ($\sim$ 9000~e$^-$) is
  transported to the readout electronics on a set of 55~cm long, 5-layer, Kapton flexible printed circuit boards~\cite{kadflx} with a
  capacitance of 60-90~pF. The noise and gain for a typical channel was $\sim$ 1000~e$^-$ and $\sim$
  100~mV/fC respectively. The low backgrounds resulting from the insensitivity of the diamond detectors to
  synchrotron radiation, together
  with the low noise of the readout, in spite of its large separation from the detectors, helped mitigate the challenges posed
  by the small signal size of diamond detectors.

The detectors were operated in single electron mode. The data acquisition (DAQ) system employed a set of field programmable gate array (FPGA) based logic modules~\cite{v1495}
to implement a track-finding algorithm, which generated a trigger when a strip in the same cluster of 4 adjacent strips was identified in
multiple active planes. The Compton scattered electrons are approximately perpendicular to the detector planes and almost co-linear with the incident electron beam, hence, they deviated by $<$ 2 strips between the planes furthest apart.  
Three detector planes were used  during the experiment, and the typical trigger condition required 2 out of 3
planes with a trigger rate of 70 - 90 kHz. The strip hits were histogrammed on the FPGA modules and read out during each helicity reversal (beam helicity was reversed at a rate of 960~Hz). Untriggered hits were also recorded and
were used for studying DAQ dead-time and trigger inefficiencies. With the track finding trigger, electronic noise was suppressed by a factor of 100~--~200 compared to the untriggered mode, which led to a significantly better signal-to-background ratio in the triggered mode,
but at the cost of a few percent DAQ inefficiency due to the combination of dead time and trigger inefficiency. Improvements
  in the DAQ design can readily eliminate these inefficiencies in the future. Although it was not needed for this applications,
  the hits on all planes and the track information can be readily used to improve the resolution of the detector, however, a careful determination of the strip-wise efficiency for each plane would be required in order to determine the total detector efficiency.

For a beam current of 180~$\mu$A and a laser intensity of 1.7~kW, the total untriggered rate in the detector was 130~--~180~kHz ($\sim$ 2.5 kHz per strip). The well
tuned electron beam, low-noise
electronics and the insensitivity of diamond to synchrotron radiation contributed to a signal-to-background 
ratio of $\mathcal{O}$(10), as demonstrated by the
  Compton and the background spectra shown in Fig.~\ref{fig:measured} (top panel). The detector efficiency was estimated to
be 70\% by comparing the expected to the observed rates. The small signal sizes, large distance
between the detector and the readout electronics, and a threshold to reduce noise led to the inefficiency.
Over the 2 year running period of the \qweak experiment, the detectors were exposed to a
radiation dose of 100~kGy from electrons (synchrotron radiation not included).  No degradation of the
detector performance was observed, demonstrating the intended radiation hardness and the stability of the charge
  collection process over extended periods, which is relevant for use of diamond as tracking detectors. The strip-to-strip variation in
  efficiency, which was shown to have negligible impact on the asymmetry measurement, may be of concern in other applications.

\section{Analysis and Results}
\label{sec:analysis}

\begin{figure}[hbtp!]
\includegraphics*[width=8.5cm]{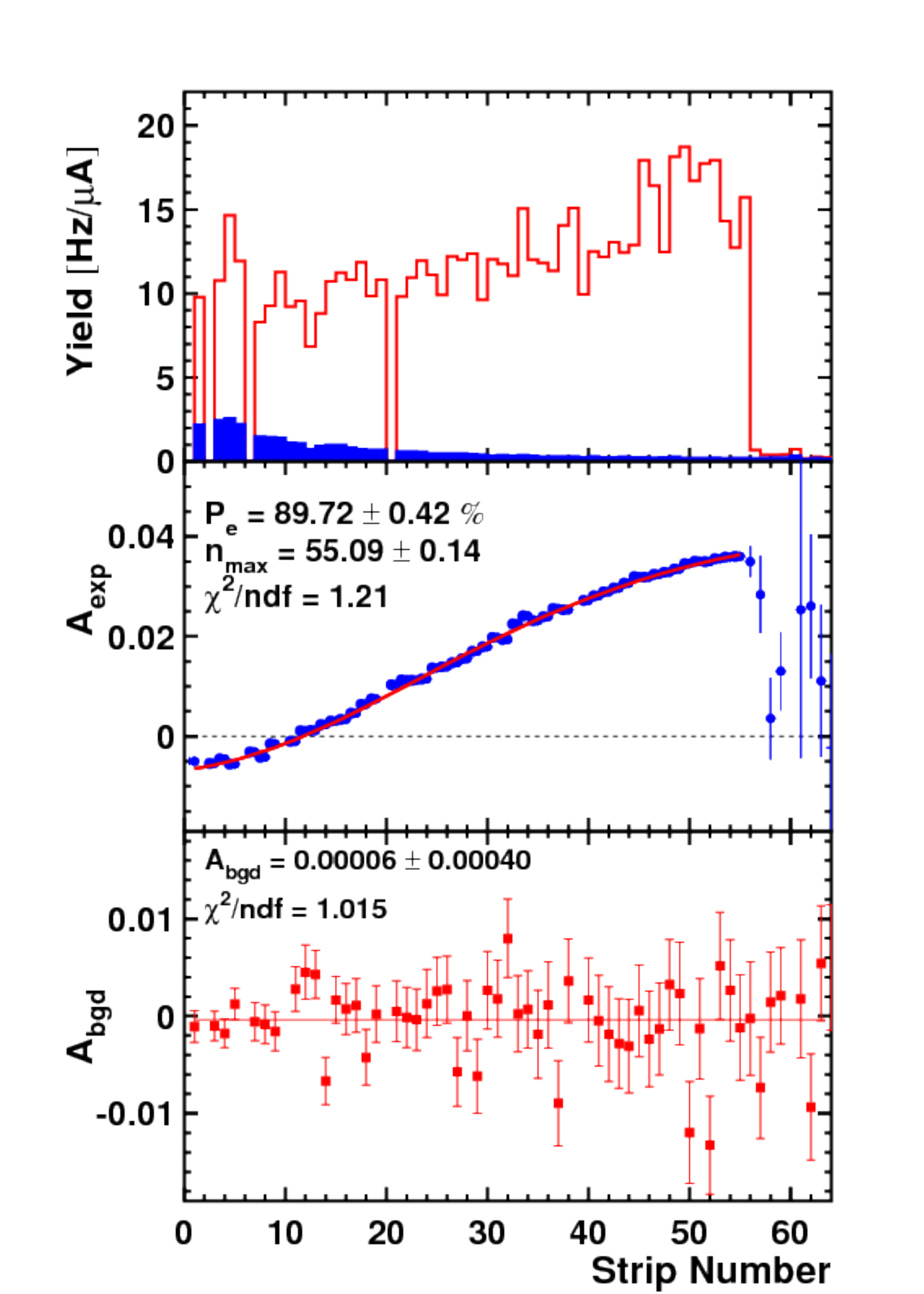}
\caption[]{Yield and asymmetry data from a single detector plane plotted versus detector strip number, for a typical hour-long run.  Statistical uncertainties only.
(top) The charge normalized yield at a beam current of 180~$\mu$A and laser intensity of 1.7~kW. 
 The laser-on yield is shown in red and laser-off (background) yield is shown in shaded blue. 
(middle) The measured Compton asymmetry (background-subtracted).  The solid red line is a fit to Eq.~\ref{eq:pol}.
(bottom) The background asymmetry from the laser-off period.   The solid red line is a fit to a constant value. } 
\label{fig:measured}
\end{figure}

The electron beam helicity was reversed at a rate of 960~Hz in a pseudo-random sequence of quartets. 
The Compton laser was operated in 90 second cycles (60~s on and 30~s off).
The laser-off data were used to measure the background, which 
was subtracted from the laser-on yield for each electron helicity state
The signal-to-background ratio was 5--20, depending on the strip.
The measured asymmetry was built from the yields using, 
\begin{equation}
A_{\rm exp} = \frac{Y^{+} - Y^{-}}{Y^{+} + Y^{-}},
\label{eq:asym}
\end{equation}
 where 
 $Y^{\pm} =  {N^{\pm}_{\rm on}}/{Q^{\pm}_{\rm on}} - {N^{\pm}_{\rm off}}/{Q^{\pm}_{\rm off}}$ 
 is the charge normalized Compton yield for each detector strip, $N^{\pm}_{\rm on/off}$ is the number of detected counts,
 and $Q^{\pm}_{\rm on/off}$ the beam charge, accumulated during the laser (on/off) period for the ($\pm$) electron helicity state. 
 A statistical precision of $<$~1\% per hour was routinely achieved. Typical yield spectra for the
 laser-on and laser-off periods are shown in Fig.~\ref{fig:measured} (top). 
Consistent results were obtained subtracting the background over 1 laser cycle (90~s) and also over $\sim$900~s.
A typical spectrum for an hour long run is shown in Fig.~\ref{fig:measured}. 
The background asymmetry is consistent with zero within the statistical uncertainties.

The electron beam polarization $P_e$ was extracted by fitting the measured asymmetry to the theoretical Compton asymmetry, using
\begin{equation}
A_{\rm exp}^n = P_eP_{\gamma}A_{\rm th}^n,
\label{eq:pol}
\end{equation}
where $P_{\gamma}$ is the polarization of the photon beam and $A_{\rm th}^n$ is the $\mathcal{O}(\alpha)$ theoretical Compton  asymmetry
for fully polarized electrons and photon beams in the $n$-th strip.  
The theoretical Compton  asymmetry $A_{\rm th}(\rho)$ was calculated as a function of the dimensionless variable
\begin{equation}
\rho=\frac{E_{\gamma}}{E_{\gamma}^{\rm max}} \approx \frac{E_e^{\rm beam}-E_e}{E_e^{\rm beam}-E_e^{\rm min}},
\label{eq:rho}
\end{equation}
where $E_{\gamma}$ the energy of a back-scattered photon,  $E_{\gamma}^{\rm max}$ is the maximum allowed photon energy, and 
$E_e$, $E_e^{\rm min}$, and $E_e^{\rm beam}$ are the scattered electron energy, its minimum value, and the
electron beam energy, respectively.
$A_{\rm th}^n$ is related to $A_{\rm th}(\rho)$  by mapping $\rho$ to the strip number. The mapping is performed using knowledge of the magnetic field in the third dipole, 
the geometry of the chicane, the strip pitch and the position of the kinematic
end point (Compton edge) expressed as a strip position, $n_{\rm max}$.
An initial estimate of the kinematic end-point, $n_{\rm max}$, was determined from the edge of the yield spectrum. It was observed to vary
slowly,
as the electron beam angle drifted, by up to $\pm$0.5~mrad.

Radiative corrections to the Compton asymmetry were calculated to leading order with a low energy approximation applicable for few GeV
electrons~\cite{radcor2}.  
The radiative correction to the asymmetry was $<$0.3\% in all strips.

Equation~\ref{eq:pol} was fit to the measured asymmetries with $P_e$ and $n_{\rm max}$ 
as the two free parameters.  
No systematic deviation of the shape of the asymmetry was observed.  
A typical fit is shown in Fig.~\ref{fig:measured}. The $\chi^2$ per degree-of-freedom of the fit, considering statistical uncertainties only,
ranges between 0.8 and 1.5 for 50~--~60 degrees of freedom. 
The detection of a large fraction of the Compton electron spectrum, spanning both sides of the zero crossing of the Compton asymmetry,
significantly improved the robustness of the fit. The fit quality was validated using the simulation discussed below.

The systematic uncertainty in the determination of $P_e$ is summarized in Table~\ref{tab:tab1}.   In previous polarimeters using a laser system based
on a \fp cavity, knowledge of the laser polarization was a significant source of uncertainty.  
Previous quoted uncertainties for the laser polarization have been larger than the total uncertainty for the present measurement. At Jefferson Lab, for example 
  uncertainties ranging from 0.6\% to 1.1\% have been reported~\cite{ HallA, Escoffier, Friend} 

Pressure induced birefringence in the vacuum
window can lead to large changes in laser polarization which cannot be directly measured in the evacuated beamline. More recently, a precision
  of 0.3\%~\cite{LPOL2000} has been quoted for laser polarization in a \fp cavity, however even in this case, the birefrigence in the vacuum window was not measured directly.  Our experience suggests that without direct measurements of the window birefrigence, inferred knowledge of the laser polarization in the \fp cavity can be flawed. A technique that bypasses this requirement is needed.

\begin{figure}[hbtp!]
\includegraphics*[width=8.5cm]{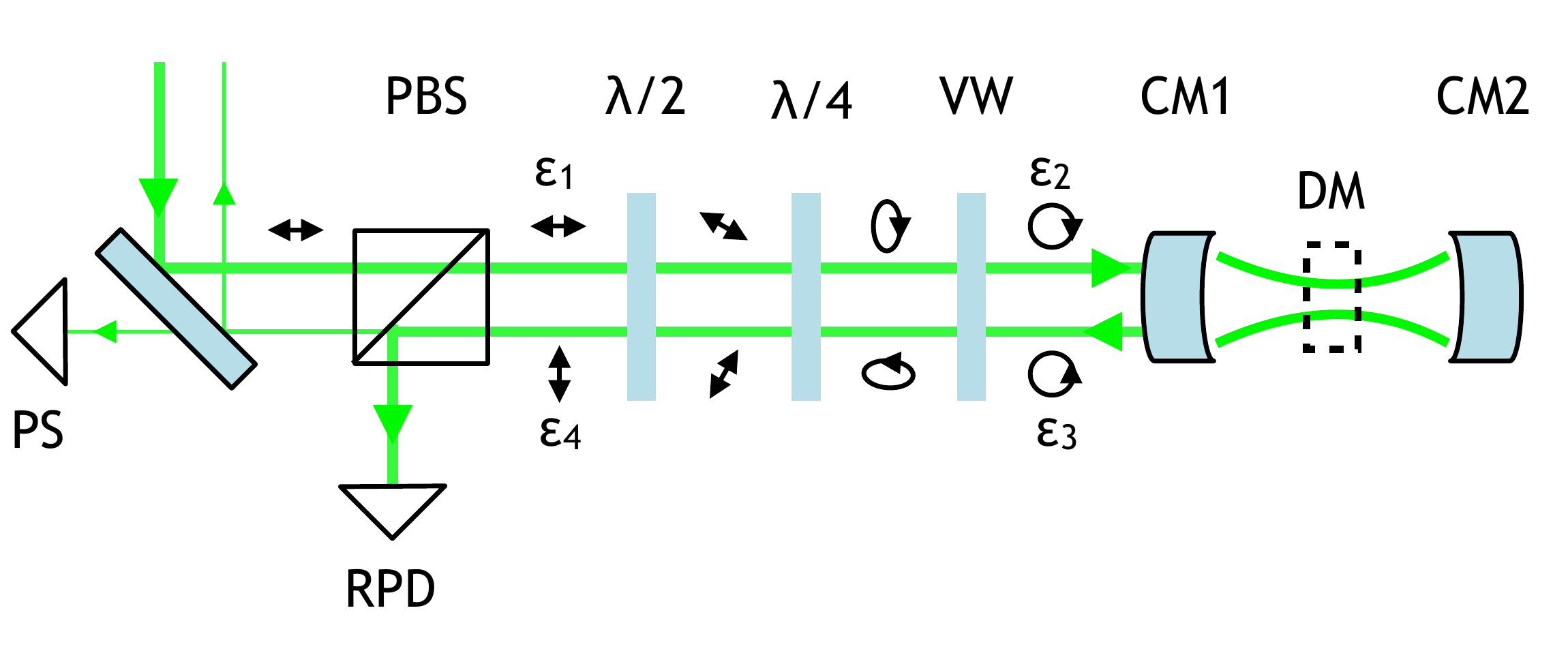}
\caption[]{Scheme for maximizing the circular polarization at the cavity.  Laser light entering the system passes through a polarizing beam splitter (PBS), half-wave plate ($\lambda/2$), quarter-wave plate ($\lambda/4$), and vacuum window (VW) before it is either reflected off the cavity entrance mirror (CM1) or becomes resonant in the cavity.  Note that in this figure, the element VW also includes 3 steering mirrors, which are incorporated in the model but left out of the figure for simplicity. Depending on the polarization state at CM1, reflected light will either arrive in reflected photo-diode (RPD), used for frequency-locking feedback, or be sampled by the polarization signal (PS) photodiode behind a steering mirror.  Light arriving at the cavity entrance mirror is fully circular when there is no signal in PS. Before the experiment, with part of the beamline vacuum pipe removed, it was possible to do a direct measurement (DM) of the circular polarization in the cavity.}
\label{fig:laser_schema}
\end{figure}

Figure~\ref{fig:laser_schema} shows our implementation of a scheme based on an optical reversibility theorem~\cite{reverse}, which relates the polarization ellipticity at the output of an optical system to the polarization of the retro-reflected light at the input, in order to maximize the circular polarization in the cavity. The technique works by analyzing the light reflected from the entrance mirror of the cavity. A polarizing beam splitter (PBS), half-wave plate, and quarter-wave plate were used to create an arbitrary polarization state which was then propagated to the cavity through an optical system with unknown birefringence, dominated by a vacuum window.  Minimizing the polarization signal, the back-reflected light that is transmitted through the PBS, maximizes the degree-of-circular-polarization (DOCP) at the cavity entrance mirror. 

Representing the  initial (linear) laser polarization state after the polarizing beam splitter (PBS)  as ${\bf{\varepsilon_1}}$, the polarization vector at the first cavity mirror (${\bf{\varepsilon_2}}$) is given by ${\bf{\varepsilon_2}} = M_E{\bf{\varepsilon_1}}$. Here $M_E$ represents the transport through all the optical elements to the cavity entrance mirror ($\lambda/2$, $\lambda/4$, and VW in Figure~4). Assuming no polarization loss, transport backwards through the same optical system can be written as the transpose of the forward matrix, $M_E^T$. In the formalism of~\cite{reverse}, the polarization vector does not change when the light direction changes and the vector representing the light reflected from the cavity mirror (${\bf{\varepsilon_3}}$) is equivalent to ${\bf{\varepsilon_2}}$. Therefore, the polarization of the light reflected from the cavity after transport backwards through the optical system is, ${\bf{\varepsilon_4}} = M_E^TM_E{\bf{\varepsilon_1}}$. The optical reversibility theorem dictates that for a linear polarization vector
  ${\bf{\varepsilon_1}}$, the polarization at the cavity entrance (${\bf{\varepsilon_2}}$) is circular only if
  ${\bf{\varepsilon_4}}$ is linear and orthogonal to ${\bf{\varepsilon_1}}$. This means that the PBS, which creates the initial linear polarization state will not allow the orthogonal reflected state to pass through- hence minimization of the laser power propagating backwards through this cube ensures circular polarization at the cavity. Ref.~\cite{LPOL2000} used an implementation of the optical reversibility theorem that is equivalent to maximizing the signal in the RPD of Fig.~\ref{fig:laser_schema}.  This is less sensitive than measuring polarization in extinction, such as minimizing the signal in the PS of Fig.~\ref{fig:laser_schema}, and was not used as their primary method.

\begin{figure}[hbtp!]
\includegraphics*[width=8.5cm]{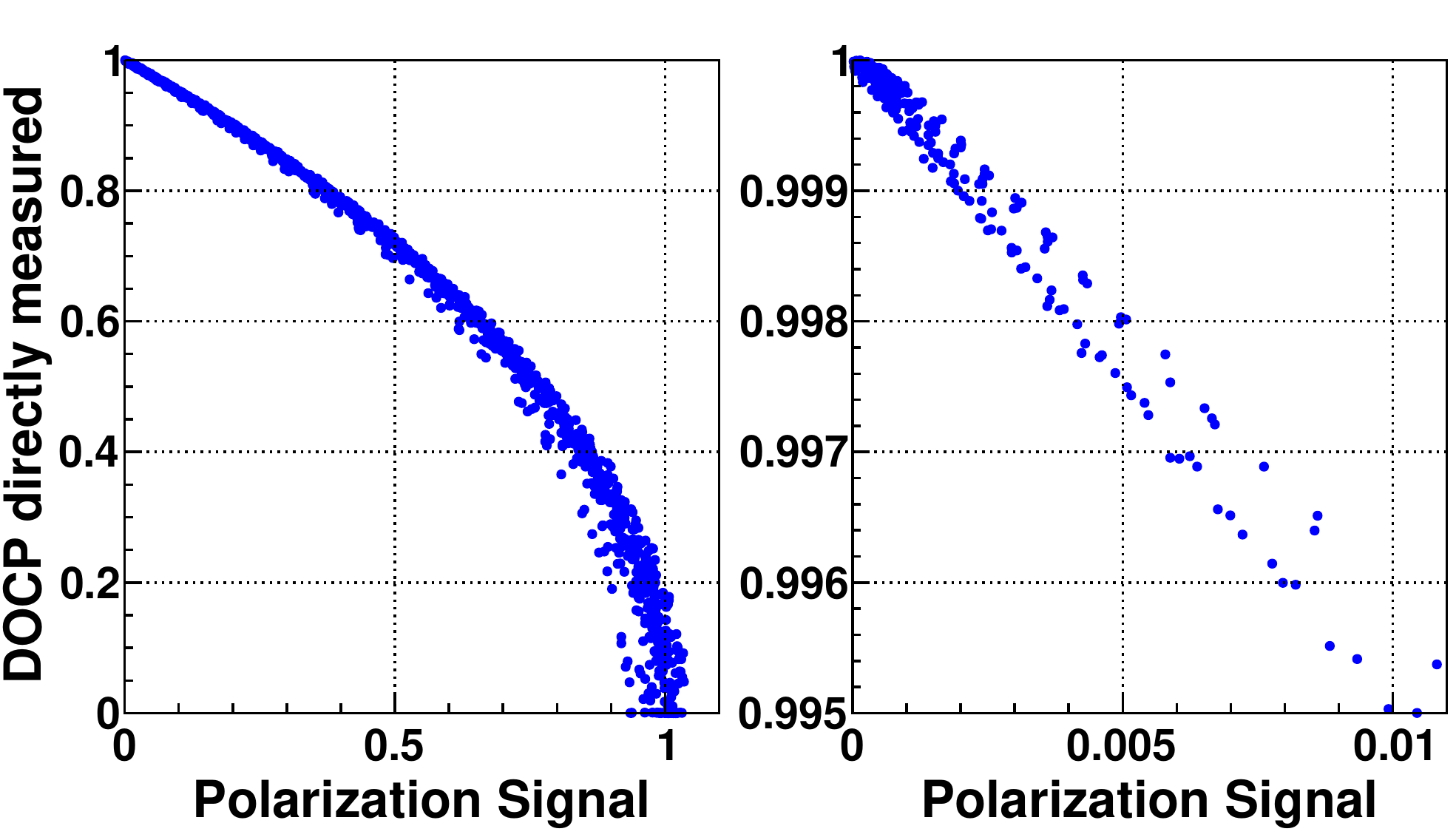}
\caption[]{Direct test of laser polarization maximization technique. The correlation between the laser DOCP directly measured after the cavity mirror and the Polarization Signal extracted from the reflected light. The left panel shows the full range while the right panel is a
  zoom of the region of maximum DOCP.}
\label{fig:laser_directcheck}
\end{figure}

To determine the uncertainty in the photon polarization, this DOCP maximization technique was 
directly tested {\it in situ}.
With the vacuum enclosure removed, the intra-cavity DOCP was measured simultaneously with the polarization signal while scanning over
input polarization states, with a concentration of points near the maximum DOCP, as in Fig.~\ref{fig:laser_directcheck}, demonstrating a very close and robust
correlation.  The uncertainty on the laser polarization is estimated to be 0.18\%, which is dominated by our ability to bound, through
direct measurement, effects that might alter the polarization over the numerous reflections within the \fp cavity. It is expected that this bound can be improved following methods implemented in Ref.~\cite{Asenbaum}, in particular using an optical isolator to capture the full polarization signal.  This would improve the signal to background and allow studying a locked cavity with arbitrary polarization (only near-circular polarization is possible with the current system.) Effects of analyzing power, depolarization, or spatial polarization gradients are bound by the degree of extinction in the Polarization Signal, and are included in the quoted laser polarization uncertainty~\cite{jonesthesis}.

\begin{figure}[hbtp!]
\includegraphics*[width=8.0cm]{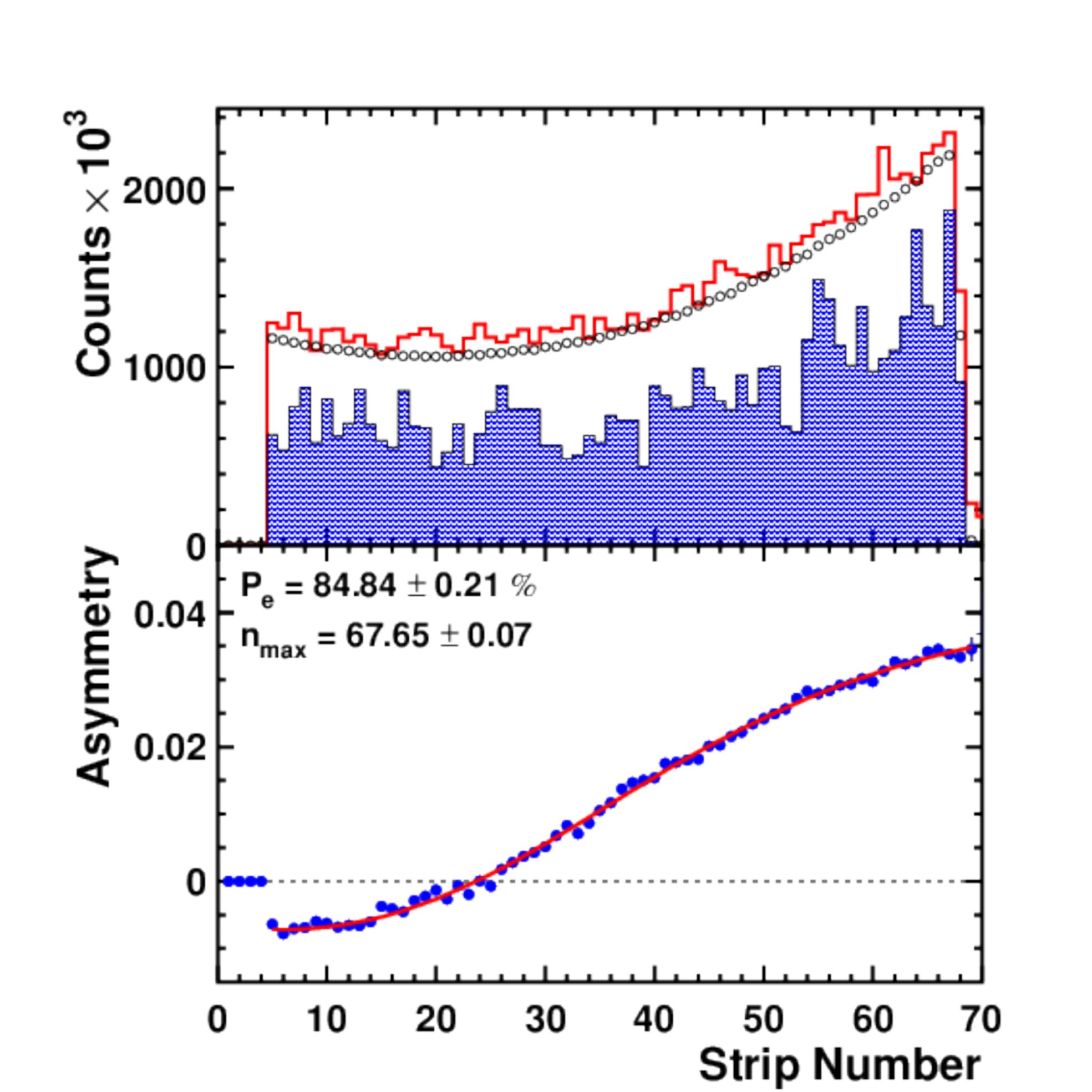}
\caption[]{(top) Typical Monte Carlo simulated Compton spectra for a single detector plane; ideal (black open circles), with noise (red) and
  with detector efficiency (blue, shaded). (bottom) The Compton asymmetry extracted from the simulated spectrum including detector
  efficiency (blue circles), and a two parameter fit to the calculated asymmetry (red line). The input asymmetry was 85\%.} 
\label{fig:mc}
\end{figure}

The uncertainties in the measured asymmetry were studied using a Monte Carlo simulation of the Compton polarimeter, which was coded using the
GEANT3~\cite{g3} detector simulation package. In addition to Compton scattering, the simulation included backgrounds from beam-gas
interactions and beam halo interactions in the chicane elements. It also incorporated the effects of detector efficiency,
the track-finding trigger, and electronic noise. A typical simulated strip-hit spectrum (ideal, with noise, and with noise and efficiency),
and the asymmetry extracted from it, are shown in Fig.~\ref{fig:mc}. 
The simulation was used to study the analysis procedure and the statistical quality of the fits that were used to extract the beam polarization.
It was
demonstrated that the central value of the polarization fit parameter was typically insensitive to small distortions to the electron
spectrum such as a few missing or noisy strips, and the observed strip-to-strip variation in efficiency. The simulation was also used to
study a variety of sources of systematic uncertainties.  
For each source, the relevant parameter was varied within the expected range of uncertainty, and the range of variation of the
extracted polarization was listed as its contribution to the systematic uncertainty.

The Monte Carlo simulation demonstrated that secondary particles knocked out by the Compton 
scattered electron passing through the first detector plane produced a 0.4\% change in the extracted polarization
in the subsequent planes, consistent with observation. A correction for the second and third planes could be made but at the cost of a
slightly higher systematic uncertainty and hence only the results from the first detector plane are quoted here. Although all three planes
were used in the tracking trigger, the results from the first detector plane were shown by the simulation to be insensitive to this effect.

There were several sources of rate-dependent efficiency associated with the DAQ system, such as the algorithm used to identify
electron tracks and form the trigger, and the dead-time due to a busy (hold off) period in the DAQ. 
A digital logic simulation platform, Modelsim~\cite{modelsim}, was used to model the DAQ system. Simulated Compton events, backgrounds,
and noise signals were processed with this model, which made a detailed accounting of the logic and delays from the internal signal
pathways in the FPGA modules and the external electronic chain.

These results were used to determine a correction to the detector yields, for each hour-long run, based on the detector rates during the
run.  This correction is calculated and applied for each beam helicity state independently. An estimate of the systematic uncertainty due
to this correction was determined from the variation of the ratio of the polarizations extracted from the corrected, triggered data to those
obtained from the untriggered data over a wide range of signal rates and several difference trigger conditions. 
The DAQ efficiency correction resulted in $<$~1\% change in the extracted polarization.

\setlength{\tabcolsep}{4pt}
\begin{table}[hbtp!]
\caption{Systematic Uncertainties}
\label{tab:tab1}
\begin{center}
{\small 
\begin{tabular}{|l|c|c|}
\hline
\textbf{Source} & \textbf{Uncertainty} & \textbf{$\Delta$P/P\%} \\
\hline \hline
\textbf{Laser Polarization}              & \textbf{0.18\%}          &      \textbf{0.18} \\
\hline\hline

helicity correl. beam  & 5 nm, 3 nrad & $<$ 0.07 \\
Plane to Plane                  &  secondaries  &       0.00    \\
magnetic field                  &  0.0011 T             &       0.13    \\
beam energy                     &  1 MeV                &       0.08    \\
detector z position     &  1 mm                 &       0.03    \\
trigger multiplicity             &  1-3 plane    &       0.19    \\
trigger clustering              &  1-8 strips   &       0.01    \\
detector tilt (x, y and z) &  1 degree     &       0.06    \\
detector efficiency             &  0.0 - 1.0    &       0.1     \\
detector noise          & up to 20\% of rate   &       0.1     \\
fringe field            & 100\%                 &       0.05    \\
radiative corrections   & 20\%                  &      0.05    \\
DAQ efficiency correction & 40\%  &    0.3     \\
DAQ efficiency pt.-to-pt. &      &   0.3 \\
Beam vert. pos. variation & 0.5 mrad & 0.2 \\
spin precession in chicane & 20 mrad & $<$ 0.03 \\ 
\hline
\textbf{Electron Detector Total} & & \textbf{0.56} \\
\hline \hline
\textbf{Grand Total}   &                               & \textbf{0.59} \\
\hline

\end{tabular}
}
\end{center}
\end{table}

\begin{figure}[hbtp!]
\includegraphics*[width=8.0cm]{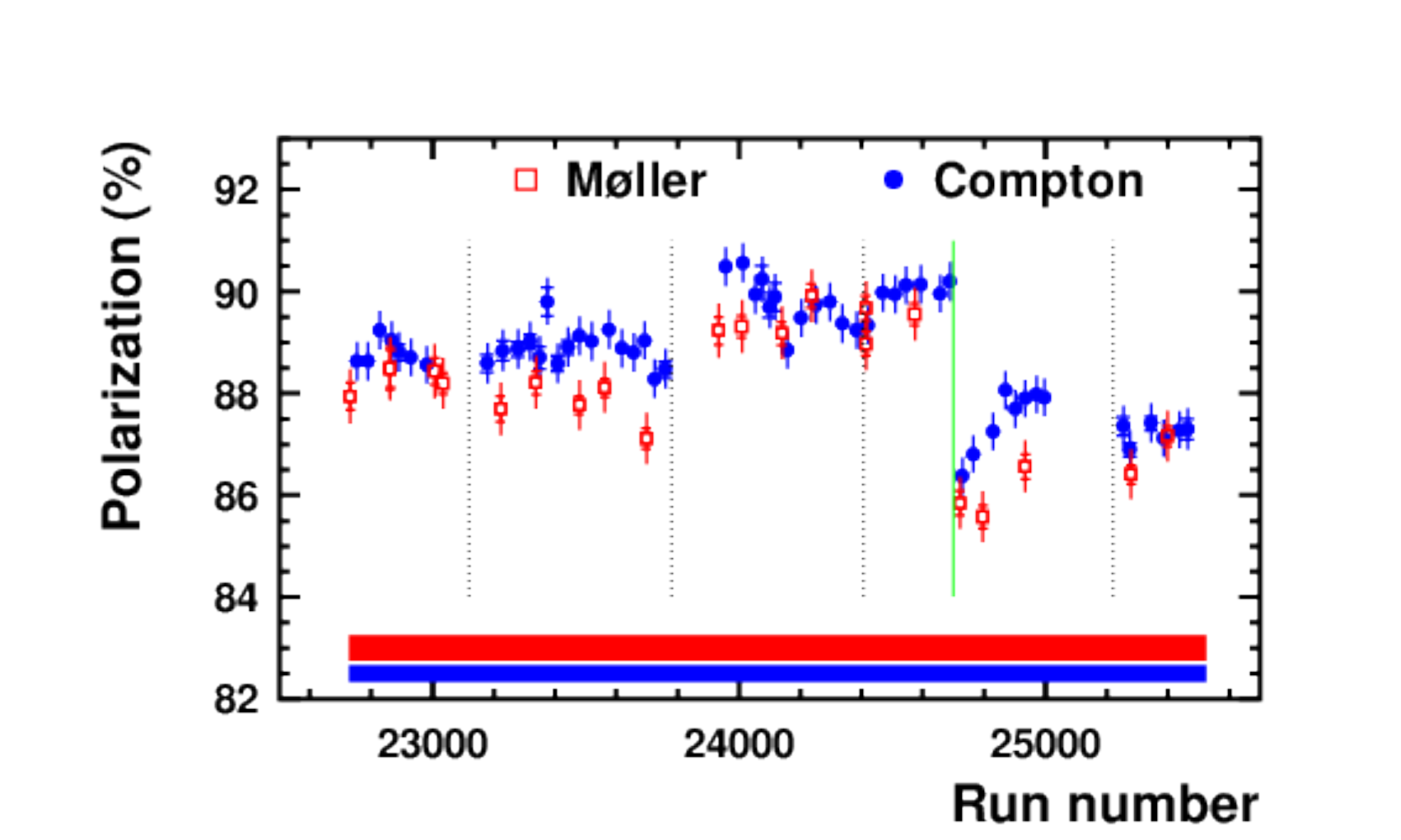}
\caption[]{The extracted beam polarization for the 1.16~GeV, 180~$\mu$A electron beam, as a function of run-number and averaged over 30~hour
  long periods, during the second 
run period of the \qweak experiment (blue, solid circle). Also shown are the results from the intermittent measurements with 
the M{\o}ller polarimeter~\cite{qweaknim} (red, open square). The inner error bars show the statistical uncertainty while 
the outer error bar is the quadrature sum of the statistical and point-to-point systematic uncertainties. The solid bands show 
the additional normalization/scale type systematic uncertainty (0.42\% Compton and 0.65\% M{\o}ller). The dashed and solid (green) vertical
lines indicate changes at the electron source.} 
\label{fig:polarization}
\end{figure}

The extracted beam polarization for the entire second running period of the \qweak experiment is shown in Fig.~\ref{fig:polarization}.
Changes at the electron source, indicated by the dashed and solid vertical lines, led to discontinuities in the beam polarization. Each
point is shown with systematic uncertainties that may vary for each measurement, while a common systematic uncertainty of 0.42\% applies
to all points together.

These results are quantitatively compared to results~\cite{qweaknim} from the M{\o}ller polarimeter by examining periods of stable
polarization between changes in the polarized source.
Previous cross-comparisons between polarimeters in this energy range have uncovered significant discrepancies between various
polarimeters~\cite{Grames}. 
The ratio of Compton to M{\o}ller measurements, when averaged over these stable periods using statistical and point-to-point systematic
uncertainties, was ~1.007~$\pm$~0.003.  
The results are compatible within the total relative normalization uncertainty of 0.77\%. 
This is the first direct comparison of two independent polarimeters with better than 1\% precision. 

Future experiments will require a polarimetry precision of 0.4\% with beam energies between 6 and 11~GeV.  Our results indicate that these
goals are within reach of Compton polarimetry.  Recent results using a photon detector in integrating mode~\cite{Friend} have demonstrated that
uncertainties in the photon analysis (excluding the laser polarization) are at the level of 0.5\%. Such a measurement could be combined
with an independent electron analysis as demonstrated here, with a precision approaching 0.5\%, with the dominant systematic error in
common between the two analyses being the uncertainty on intra-cavity laser polarization ($<0.2\%$).  
It is worth noting that further gains are possible: the dominant errors in the electron analysis relate to rate-dependent DAQ inefficiencies,
which would undoubtedly be reduced through refinement of the logic and timing parameters, while improvements in gain stability and linearity
measurements would further improve the photon measurements.  The increased beam energies for planned future measurements are also more
favorable to Compton polarimetry.

\section{Conclusions}
\label{sec:res}

The polarization of a 1.16~GeV CW electron beam was measured with a systematic uncertainty of 0.59\%. 
The interacting photon polarization was maximized and the uncertainty reduced using a novel technique based on the reflected incident
light. We used diamond microstrip detectors for the first time as tracking detectors and demonstrated their ability to withstand a high
radiation dose and their stability over long periods. The high granularity of the detectors and the measurement of a large fraction of the Compton electron spectrum, spanning the asymmetry zero crossing, coupled with a robust analysis technique and rigorous simulations of the polarimeter and the DAQ system, produced a reliable, high precision measurement of the polarization in a high radiation environment. Due to these technical advances, the uncertainty goal was significantly surpassed. These results suggest that even more precise electron beam polarization measurements, such as required for the future parity-violation measurements SOLID and MOLLER,  will be achievable through Compton polarimetry. Further, diamond-based tracking detectors are the superior choice for high radiation environments and should find more wide-spread use.

\section{Acknowledgments}
This work was funded by the U.S. Department of Energy, including contract 
\#AC05-06OR23177 under which Jefferson Science Associates, LLC operates Thomas Jefferson National Accelerator Facility, and by the U.S. National Science Foundation and  the Natural Sciences and Engineering Research Council of Canada (NSERC). We wish to thank the staff of JLab, TRIUMF, and MIT-Bates, for their vital support. We thank H. Kagan from Ohio State University for teaching us about diamonds, training us on
characterizing them and helping us build the prototype detector, and Tanja Horn and Ben Raydo at Jefferson Lab for suggesting and assisting with the FPGA-based readout scheme for the electron detector. We also acknowledge the U. of Manitoba Nano Systems Fabrication Lab for the
use of their facilities.


\begin{thebibliography}{99}
\bibitem{qweakprl}D.~Androic {\it et al.}, First Determination of the Weak Charge of the Proton, Phys. Rev. Lett. {\bf 111}, 141803 (2013). 
\bibitem{qweaknim}T. Allison {\it et al.}, The Q$_{weak}$ experimental apparatus, Nucl. Instr. Meth. {\bf A781},105 (2015).
\bibitem{MOLLERExp} The MOLLER Collaboration, The MOLLER Experiment: An Ultra-Precise Measurement of the Weak Mixing Angle
  Using M{\o}ller Scattering, arXiv:1411.4088.
\bibitem{SOLID}P.~Souder {\it et al.}, Precision Measurement of Parity-Violation in Deep Inelastic Scattering Over a Broad Kinematic Range, JLab Proposal {\bf PR12-10-007} (2012)
\bibitem{narayanthesis}A. Narayan, Determination of electron beam polarization using electron detector in Compton polarimeter with less than 1\% statistical and systematic uncertainty Ph. D. Thesis, Mississippi State University, 2015 (unpublished).
\bibitem{moller}M.~Hauger  {\it et al.}, A high-precision polarimeter, Nucl.\ Instrum.\ Meth.\ A {\bf 462}, 382 (2001)
\bibitem{Magee}J.~Magee, Sub-percent precision M{\o}ller polarimetry in experimental Hall C, Proc. of Sci. PSTP2013, 039 (2013).  
\bibitem{SPEAR} D. Gustavson {\it et al.}, A backscattered laser polarimeter e$^{+}$e$^{−}$ storage rings, Nucl. Instr. Meth. {\bf 165}, 177 (1979); 
\bibitem{LEP} L. Knudsen {\it et al.}, First observation of transverse beam polarization in LEP, Phys. Lett. {\bf B270}, 97 (1991).
\bibitem{Hera}D. P. Barber {\it et al.}, The HERA polarimeter and the first observation of electron spin polarization at HERA, Nucl. Instr. Meth. {\bf A329}, 79 (1993). 
\bibitem{Nikhef}I. Passchier {\it et al.}, A Compton backscattering polarimeter for measuring longitudinal electron polarization, Nucl. Instr. Meth. {\bf A414}, 446 (1998).
\bibitem{LPOL}M.~Beckmann {\it et al.}, The Longitudinal Polarimeter at HERA, Nucl.\ Instrum.\ Meth.\ A {\bf 479}, 334 (2002).  
\bibitem{Bates} W. Franklin {\it et al.}, The MIT-Bates Compton Polarimeter, AIP Conf.Proc. 675, 1058 (2003).
\bibitem{HallA} M. Baylac {\it et al.}, First electron beam polarization measurements with a Compton polarimeter at Jefferson Laboratory, Phys. Lett. {\bf B539}, 8 (2002); N. Felletto {\it et al.}, Nucl. Instr. Meth. {\bf A459}, 412 (2001).
\bibitem{Escoffier}S. Escoffier {\it et al.}, Accurate measurement of the electron beam polarization in JLab Hall A using Compton polarimetry, Nucl. Instrum. Meth. A {\bf 551}, 563 (2005).
\bibitem{Friend} M. Friend {\it et al.}, Upgraded photon calorimeter with integrating readout for the Hall A Compton polarimeter at Jefferson Lab, Nucl. Instr. Meth. {\bf A676}, 96 (2012).
\bibitem{SLD} K.~Abe {\it et al.}, High-Precision Measurement of the Left-Right Z Boson Cross-Section Asymmetry, Phys.\ Rev.\ Lett.\  {\bf 84}, 5945 (2000); M. Woods, hep-ex/9611005 (1996).
\bibitem{happex2}  A.~Acha {\it et al.}  [HAPPEX Collab.], Precision Measurements of the Nucleon Strange Form Factors at $Q^2$∼0.1 GeV$^2$, Phys.\ Rev.\ Lett.\  {\bf 98}, 032301 (2007).
\bibitem{Parno} D.S. Parno {\it et al.}, Comparison of modeled and measured performance of a GSO crystal as gamma detector, Nucl. Instr. Meth. {\bf A728}, 92 (2013).
\bibitem{PREX}   S.~Abrahamyan {\it et al.}, Measurement of the Neutron Radius of $^{208}$Pb through Parity Violation in Electron Scattering, Phys.\ Rev.\ Lett.\  {\bf 108}, 112502 (2012).
\bibitem{Bauer1995} C.~Bauer {\it et al.}, Radiation hardness studies of CVD diamond detectors, Nucl. Instrum. Methods {\bf 367}, 207 (1995).
\bibitem{Zoeller1997} M. M. Zoeller {\it et al.}, Performance of CVD diamond microstrip detectors under particle irradiation, IEEE Trans. Nucl. Sci. {\bf 44} 815 (1997).
\bibitem{Borchelt}F.~Borchelt  {\it et al.}, First measurements with a diamond microstrip detector, Nucl. Instr. Meth. A {\bf 354}, 318 (1995).
\bibitem{Tapper2000}R.~J.~Tapper, Diamond detectors in particle physics, Rep. Prog. Phys. {\bf 63}, 1273 (2000).
\bibitem{tesla}J.~Bol {\it et al.}, Beam monitors for TESLA based on diamond strip detectors, IEEE Trans. Nucl. Sci. {\bf 51}, 2999 (2004).
\bibitem{babar}M.~Bruinsma {\it et al.}, Radiation monitoring with CVD diamonds and PIN diodes at BaBar, Nucl. Instrum. Meth. A {\bf 583}, 162 (2007).
\bibitem{cdf}P. Dong, R. Eusebi, C. Schrupp, A. Sfyrla, R. Tesarek, and R. Wallny {\it et al.}, Beam condition monitoring with diamonds at CDF, IEEE Trans. Nucl. Sci. {\bf 55}, 328 (2008).
\bibitem{cms}A. Bell {\it et al.}, Fast beam conditions monitor BCM1F for the CMS experiment, Nucl. Instrum. Meth. A {\bf 614},433 (2010). 
\bibitem{gsi}J.~Pietraszko {\it et al.}, Diamonds as timing detectors for minimum-ionizing particles: The HADES proton-beam monitor and START signal detectors for time of flight measurements, Nucl. Instrum. Meth. A {\bf 618}, 121 (2010).  
\bibitem{atlas}D.~Dobos {\it et al.}, Commissioning and first operation of the pCVD diamond ATLAS beam conditions monitor, Nucl. Instrum. Meth. A {\bf 623}, 405 (2010).  

%
\bibitem{verdi} Operators Manual Verdi TM V-8/V-10 Diode-Pumped
Lasers. Coherent, Inc. 08/2005, Part No. 0174-929-00,
\bibitem{ddet} The CERN grade diamond plates were procured from Element Six, 35 West 45th St., New York, NY 10036, USA.
\bibitem{kadflx} Flexible printed circuit boards manufactured by Kadflx Inc. 74 Northeastern Blvd. Nashua, NH 03062, USA.  
\bibitem{qwad}\qweak Amplifier Discriminator (QWAD), custom built by TRIUMF, Canada.   
\bibitem{v1495} V1495 modules from CAEN Technologies, Inc. 1140 Bay Street, Suite 2C, Staten Island, NY 10305, USA
\bibitem{radcor2} A.~Denner and S.~Dittmaier, Complete $\mathcal{O}(\alpha)$ QED corrections to polarized Compton scattering, Nucl. Phys. {\bf B540}, 58 (1999).
\bibitem{LPOL2000}V.~Brisson {\it et al.}, Per Mill Level Control of the Circular Polarisation of the Laser Beam for a Fabry-Perot Cavity Polarimeter at HERA, JINST {\bf 5}, P06006 (2010).
\bibitem{reverse} N. Vansteenkiste {\it et al.}, Optical reversibility theorems for polarization: application to remote control of polarization, J.  of the Opt. Soc. of Amer. A 10 (1993) 2240.
\bibitem{Asenbaum} P.~Asenbaum and M.~Arndt, Cavity stabilization using the weak intrinsic birefringence of dielectric mirrors, 
Optics Lett. {\bf 36}, 3720 (2011)
\bibitem{jonesthesis}D.C. Jones, Measuring the Weak Charge of the Proton via Elastic Electron-Proton Scattering, Ph. D. Thesis, University of Virginia, 2015 (unpublished).
\bibitem{g3} CERN Program Library Long Write-up W5013, Unpublished (1993).
\bibitem{modelsim}  Modelsim Reference Manual, Mentor Graphics Corp., Unpublished (2010).
\bibitem{Grames}J.~Grames {\it et al.}, Unique electron polarimeter analyzing power comparison and precision spin-based energy measurement, Phys. Rev. ST Accel. Beams 7, 042802 (2004).
\end{thebibliography}
\end{document}